\documentclass[a4paper]{jpconf}
\usepackage{graphicx,amssymb}
\usepackage{hyperref}

\newcommand{\arX}[1]{\href{http://arxiv.org/abs/#1}{{\it Preprint} arXiv:#1}}
\newcommand{\be}{\begin{equation}}
\newcommand{\ee}{\end{equation}}
\newcommand{\ba}{\begin{eqnarray}}
\newcommand{\ea}{\end{eqnarray}}
\def\bs{\begin{subequations}}
\def\es{\end{subequations}}
\def\a{\alpha}
\def\b{\beta}
\def\g{\gamma}

\def\e{\epsilon}

\def\s{\sigma}

\def\vp{\varphi}
\def\dpl{\delta_{\rm Pl}}
\def\cR{{\cal R}}
\def\cH{\mathcal{H}}

\def\cN{{\cal N}}
\def\cV{{\cal V}}

\def\lp{\ell_{\rm Pl}}

\begin{document}
\title{Inflationary spectra and observations\\ in loop quantum cosmology}

\author{Gianluca Calcagni}

\address{Max Planck Institute for Gravitational Physics (Albert Einstein Institute)\\
Am M\"uhlenberg 1, D-14476 Golm, Germany}

\ead{calcagni@aei.mpg.de}

\begin{abstract}
We review some recent progress in the extraction of inflationary observables in loop quantum cosmology. Inverse-volume quantum corrections induce a growth of power in the large-scale cosmological spectra and are constrained by  observations.
\end{abstract}

Observational implications of quantum gravity present a delicate issue. Corrections to the general relativistic dynamics are expected to arise in different ways. For instance, loop corrections are always present in perturbative graviton field theory, which can be captured in effective actions with higher-curvature corrections to the
Einstein--Hilbert action. The additional terms change the Newton potential as well as the cosmological dynamics. However, in currently observable regimes the curvature scale is very small, and so one expects only tiny corrections of (adimensional) comoving size at most $\lp \cH$, where $\lp$ is the Planck length and $\cH^{-1}=a/a'$ is the comoving radius of the Hubble region ($a$ is the scale factor in the flat Friedmann--Lema\^itre--Robertson--Walker, FLRW, background and primes denote derivatives with respect to conformal time $\tau$). In such cases, tests of quantum gravity are possible at best indirectly, for instance if it provides concrete and sufficiently constrained models for inflation. So far, however, models do not appear tight enough.

In background-independent frameworks such as loop quantum gravity (LQG) \cite{Rev1}, 
stronger modifications of the theory are possible since the usual covariant continuum dynamics is generalized, and entirely new effects may be contemplated. In LQG, gauge transformations as well as the dynamics are generated by constraint equations. Since the latter are modified with respect to the classical constraints, gauge transformations change and new spacetime structures become apparent. Some of these modifications arise as follows. An inverse of metric operators is needed to construct the quantized constraints, but this inverse does not exist because metric operators have discrete spectra containing zero as an eigenvalue. Certain quantization procedures allow one to construct suitable densely-defined operators which give rise to quantum corrections sensitive to the discreteness scale. 
 The modifications are controlled by the requirement that covariance not be broken but at most deformed with respect to its ordinary incarnation, leading to anomaly-free and consistent sets of equations. The resulting dynamics and gauge-invariant observables provide the basis for a cosmological analysis.

Inverse-volume corrections constitute the first example that has been consistently implemented, when they are small, in the dynamics of loop quantum cosmology (LQC \cite{lqc}). Effective dynamical equations with inverse-volume corrections are known for pure FLRW and linear perturbations in all sectors (scalar, vector and tensor) \cite{BH1,CH,BHKS,BC}. The setting is an inflationary era driven by a slowly rolling scalar field, the only difference with respect to the standard case being the presence of the quantum corrections. For instance, the background equations of motion are $\cH^2=(8\pi G/3)\a[{\vp'}^2/(2\nu)+a^2V(\vp)]$ and $\vp''+2\cH(1-\rmd\ln\nu/\rmd\ln a^2)\vp'+\nu a^2 V_{,\vp}=0$, where $V$ is the scalar field potential and
\be
\a \approx 1+\a_0\dpl\,,\qquad \nu \approx 1+\nu_0\dpl\,,\qquad
\dpl:= (a_*/a)^\s
\ee
are the inverse-volume quantum corrections. These are parametrized as an inverse power law of the scale factor. In pure mini-superspace (no inhomogeneities), the parameters $\a_0$ and $\nu_0$ are calculable from the quantization ambiguities of the Hamiltonian constraint and also depend on $\s$, a constant determined by the way an elementary holonomy cell evolves with the Hubble expansion. A natural choice is the so-called `improved quantization' scheme \cite{APS}, where $\s=6$ and $\a_0\sim O(0.1)\sim \nu_0$. In a flat or open universe, the Hamiltonian formalism is well defined only in a finite comoving volume $\cV_0$, which is chosen \emph{ad hoc}. However, the quantum corrections are of the form $\dpl \sim (\lp^3/\cV_0)^{\s/3} a^{-\s}$, and they depend explicitly on the unphysical fiducial volume (or, in other words, on the constant $a_*$ which can be rescaled arbitrarily).

Switching on inhomogeneities, the interpretation of inverse-volume corrections and the fiducial-volume problem change considerably. Once a closed constraint algebra and the full set of linearized perturbed equations are obtained, one can study the role of different parametrization schemes and a number of other issues. We mention the conservation law for the curvature perturbation, the construction of the inflationary spectra and other cosmological observables and, finally, the extraction of observational constraints from large-scale data. These aspects, analyzed in \cite{BC,BCT1,BCT2} will be presently reviewed here. Another characteristic effect of LQC quantization, holonomy corrections, should also enter the picture, but so far the closure of the constraint algebra has been obtained only in the tensor and vector sector \cite{BH1,MCBG}. Consequently, the phenomenology of holonomy effects has been limited to the structure of the primordial bounce and to the inflationary tensor signal \cite{Mie1}.

The first step is to implement first-order perturbation theory in the classical constraints. The fundamental canonical variables, the densitized triad and the Ashtekar--Barbero connection are given by a background contribution plus a linear inhomogeneous correction: $E_i^a=p\delta_i^a+\delta E_i^a$, $A_a^i=c\delta_a^i+\left(\delta\Gamma_a^i+\gamma\delta K_a^i\right)$, where indices $i=1,2,3$ and $a=1,2,3$ run over spacetime and internal (tangent) space directions, $p=|a|^2$, $c=\gamma \dot a$ classically, $\gamma$ is the Barbero--Immirzi parameter, $\Gamma$ is the spin connection and $K$  is the extrinsic curvature. Perturbations obey the canonical Poisson relation $\{\delta K_a^i({\bf x}),\delta E_j^b({\bf y})\}=8\pi G\delta_a^b\delta_j^i\delta({\bf x},{\bf y})$. The smeared effective Hamiltonian constraint with inverse-volume correction function reads $H[N]\sim \int \rmd^3 x N [\a(E)\cH_g+\nu(E)\cH_\pi+\rho(E)\cH_\nabla+\cH_V]$, where $N$ is the lapse function and different contributions $\cH_{g,\pi,\nabla,V}$  pertain the gravitational sector and the scalar field kinetic, gradient, and potential terms, respectively. Similarly, one considers the perturbed Gauss and diffeomorphism constraints, and imposes closure of the effective constraint algebra, $\{C_\a,C_\b\}=f_{\a\b}^{\ \ \g}(A,E) C_\g$. The resulting perturbed equations contain counterterms which fix the functions $\a$, $\nu$ and $\rho$ and guarantee anomaly cancellation in the constraint algebra \cite{BH1,BHKS}. These counterterms only depend on the three parameters $\a_0$, $\nu_0$ and $\s$, but a consistency condition further reduces the parameter space to two dimensions. Notably, inflationary and de Sitter background solutions exist for $0\lesssim \s\lesssim 3$ \cite{BC}, which is in tension with the mini-superspace parametrization $\s=6$. This problem is bypassed in the lattice refinement framework, where the proper fiducial volume $\cV=\cV_0a^3$ is replaced by the volume $L^3= \cV/\cN$ of a microscopic cell of size $L$, determined by the number of vertices (or `patches') of an underlying discrete state. The number of cells can be parametrized as $\cN=\cN_0a^{6n}$, where $0\leq n\leq 1/2$ (both $\cN$ and $L$ must not decrease with the volume in a discrete geometrical setting). Then, $\dpl:= (\lp/L)^m= (\lp^3\cN_0/\cV_0)^{m/3} a^{-(1-2n)m}=:(a_*/a)^\s$, where $m$ is a positive parameter. Eventually, one can argue that the parameter range is $\s\geq 0$ and $\a_0\geq 0$, $\nu_0\geq 0$ \cite{BC}.

Scalar fluctuations $\Psi$ and $\delta\vp$ in the metric and in the scalar field generate the gauge-invariant curvature perturbation on comoving hypersurfaces $\cR = \Psi+(\cH/\vp')(1-\s\nu_0\dpl/6)\delta\vp$. At large scales, this quantity is conserved thanks to a delicate cancellation of counterterms \cite{BC}: $\cR'=[1+(\a_0/2+2\nu_0)\dpl][\cH/(4\pi G \vp'^2)]\Delta\Psi$,
where $\Delta$ is the Laplacian. Because of this property, one can argue (and also rigorously show) that the Mukhanov scalar variable $u=z\cR$, where $z:=(a\vp'/\cH)[1+(\a_0/2-\nu_0)\dpl]$, obeys the simple dynamical equation $u''-(s^2\Delta+z''/z)u =0$, where $s(\a_0,\nu_0,\s)$ is the propagation speed of the perturbation. According to the inflationary paradigm, observables are expanded in terms of small slow-roll parameters $\epsilon := 1-\cH'/\cH^2$, $\eta := 1-\vp''/(\cH\vp')$, $\xi^2:= \cH^{-2}(\vp''/\vp')'+ \epsilon+\eta-1$. Asymptotic solutions to the Mukhanov equation at large scales, evaluated at horizon crossing, eventually yield the scalar spectrum, the scalar index and its running:
\ba
{\cal P}_{\rm s} &:=& \frac{k^3}{2\pi^2z^2} \left\langle |u_{k\ll\cH}|^2\right\rangle\Big|_{k|\tau|=1}= \frac{G}{\pi}\frac{\cH^2}{a^2\e}\left(1+\g_{\rm s}\dpl\right)\,,\\
n_{\rm s}-1 &:=& \frac{\rmd\ln {\cal P}_{\rm s}}{\rmd\ln k}=2\eta-4\e+\s\g_{n_{\rm s}}\dpl\,,\\
\a_{\rm s} &:=& \frac{\rmd n_{\rm s}}{\rmd\ln k}=2(5\e\eta-4\e^2-\xi^2)+\s(4\tilde\e-\s \g_{n_{\rm s}})\dpl = O(\e^2)+ O(\s\dpl)\,,
\ea
where $\g_{\rm s}$ and $\g_{n_{\rm s}}$ depend on the parameters $\a_0$, $\nu_0$, $\s$ in a precise way \cite{BC}. One can notice a large-scale enhancement of power via the term $\dpl\sim a^{-\s}\sim (1/|\tau|)^{-\s}\sim k^{-\s}$. If large enough, quantum corrections dominate and $\a_{\rm s}=\s f_{\rm s}(\a_0,\s)\dpl$, where $f_{\rm s}$ is a specific function of the parameters. Bounds on the scalar running, in fact, turn out to be the main constraint on the parameters. The power spectrum can be expressed in terms of the comoving wavenumber $k$ of perturbations and of a pivot scale $k_0$, which must be chosen within the scale range probed by the given experiment. Letting $x := \ln(k/k_0)$, one has
\be
\frac{{\cal P}_{\rm s} (k)}{{\cal P}_{\rm s} (k_0)} = \exp \left\{ [n_{\rm s}(k_0)-1]x+\frac{\alpha_{\rm s}(k_0)}{2}
x^2+f_{\rm s} \dpl(k_0) \left[x \left(1-\frac12 \sigma x \right)+\frac{1}{\sigma} (e^{-\sigma x}-1) \right] \right\}\,.
\ee
Due to cosmic variance, there is an intrinsic uncertainty in the determination of the spectrum at large scales (small multipoles $\ell$), which should be compared with the strength of the typical signal from quantum corrections. Tensor observables can be calculated analogously and display the same type of corrections. From those, one can extract a consistency relation between scalar and tensor perturbations: $r = -8\{n_{\rm t}+[n_{\rm t}(\g_{\rm t}-\g_{\rm s})+\s\g_{\rm t}]\dpl\}$. 

\begin{figure}
\caption{Left: Combined marginalized distribution for the quantum-gravity parameter
$\delta(k_0)=\a_0\dpl (k_0)$ and the slow-roll parameter $\e(k_0)$ with the pivot $k_0=0.002$ Mpc$^{-1}$
for $V\propto \vp^2$ and $\sigma=2$. Internal and external solid lines correspond to 
the 68\% and 95\% confidence levels, respectively. Right: Primordial scalar power spectrum ${\cal P}_{\rm s}(\ell)$ for the same potential and pivot wavenumber (corresponding to $\ell_0=29$), with three different values of $\delta(k_0)$: 0 (dotted line), $7\times 10^{-5}$ (experimental upper bound, solid line), $4.8\times 10^{-4}$ (from the a-priori upper bound $\dpl(\ell=2)<0.1$, dashed line). The shaded region is affected by cosmic variance \cite{BCT1,BCT2}.
\label{figure}}
\begin{center}
\includegraphics[width=6.0cm]{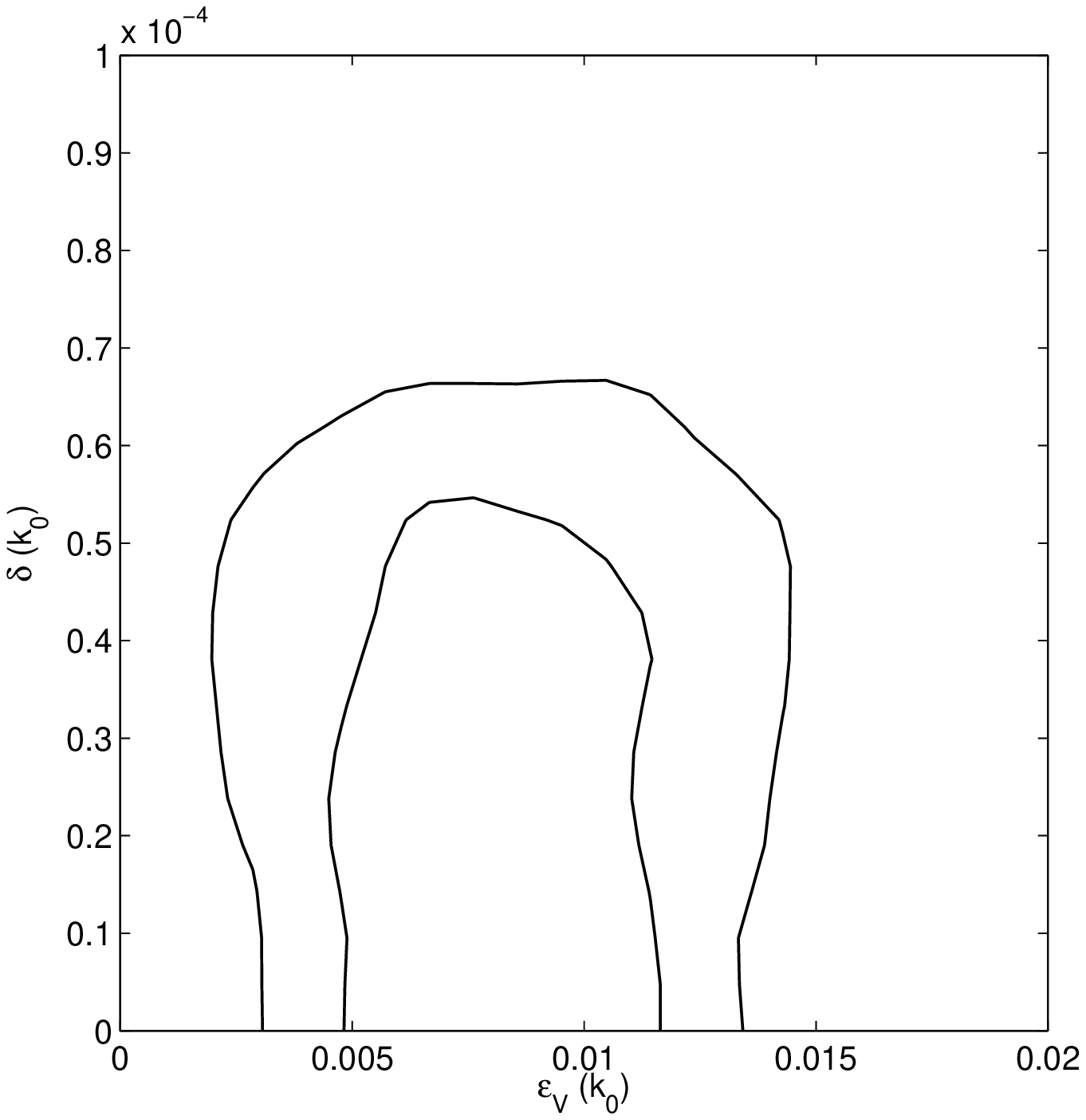}\qquad
\includegraphics[width=7.0cm]{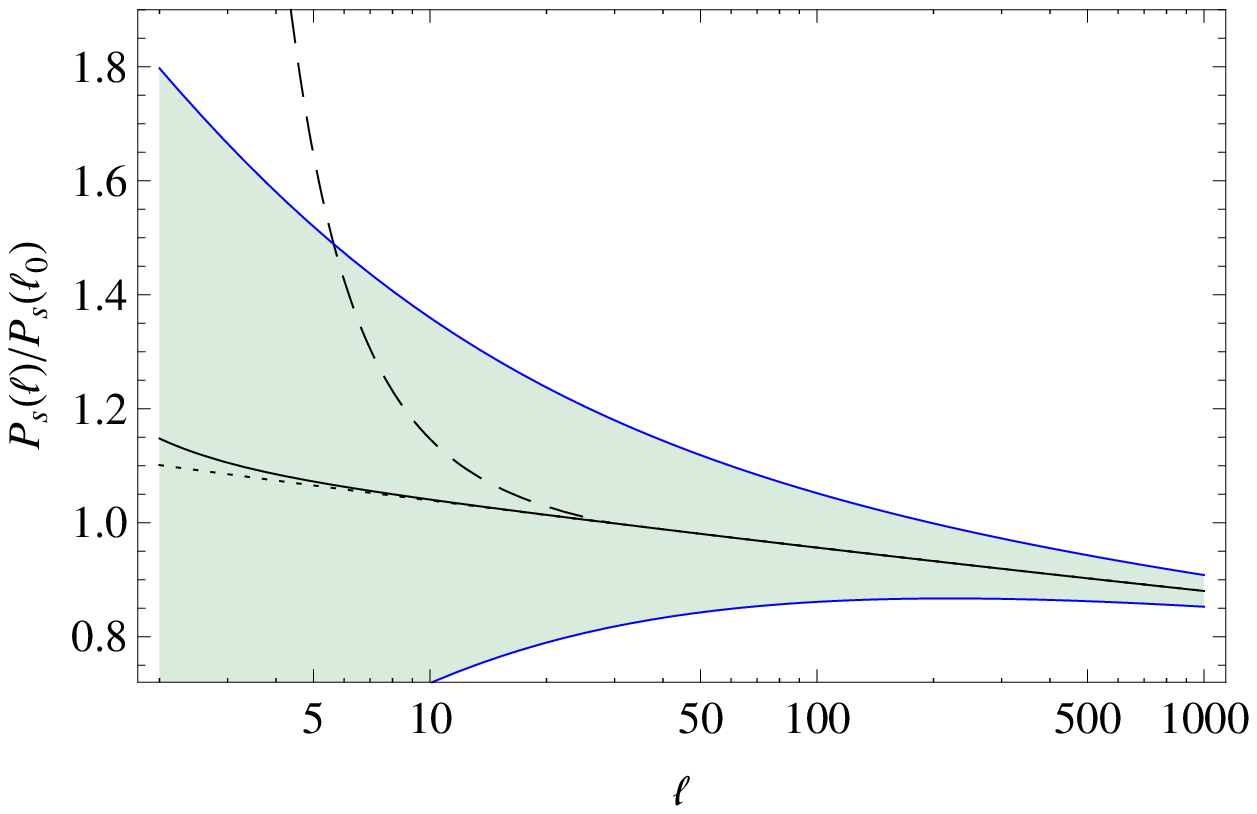}
\end{center}
\end{figure}

In order to compare with observations, one can choose an inflationary potential and recast all observables as its functions. Then, for any given choice of $\s$, one can find an upper bound for the quantum parameter $\delta:=\a_0\dpl$ (we recall that $\nu_0$ is not independent). An example of likelihood contours and power-spectrum enhancement is illustrated in figure \ref{figure}, for a quadratic potential. We use the 7-year WMAP data combined with large-scale structure, the Hubble constant measurement from the Hubble Space Telescope, Supernovae type Ia and Big Bang Nucleosynthesis.
For a range of e-folds between 45 and 65, the probability distribution of $\e$ is consistent with the theoretical range $0.008<\e<0.011$, and the quadratic potential is compatible with observations as in the standard case. In the figure, cosmic variance is greater than the LQC signal, but this is \emph{not} the case for $\sigma\lesssim 1$ \cite{BCT2}. For $\s>2$, even tighter bounds are obtained, corresponding to practically unobservable inverse-volume effects. The 95\%-confidence-level upper limits of $\delta$ constrained by observations for the potential $V(\varphi)=V_0 \varphi^2$, with $k_0=0.002~{\rm Mpc}^{-1}$ and different values of $\sigma$, are \cite{BCT2}
\ba
&&\delta(\s=0.5)=0.27\,,\qquad \delta(\s=1)=3.5 \times 10^{-2}\,,\qquad  \delta(\s=1.5)=1.7 \times 10^{-3}\,,\nonumber\\
&&\delta(\s=2)=6.8 \times 10^{-5}\,,\qquad \delta(\s=3)=4.3 \times 10^{-7}\,.
\ea
The likelihood analysis has not been performed for $\s=6$ since the signal is below the cosmic variance threshold already when $\s=2$. For $\s=3$, the parameter $\delta=\nu_0\dpl$ has been used instead. The analysis of \cite{BCT2} includes also different potentials and values of the pivot scale.

To conclude, observations highlight a tension between mini-superspace parametrizations of FLRW LQC ($\s>1$) and lattice refinement parametrizations. The latters are the only one compatible (at least in this model)
with anomaly cancellation in inhomogeneous LQC and power-law inflationary solutions. Tight upper bounds can be obtained for inverse-volume quantum corrections; their improvement with future missions such as Planck will further constrain the parameter space of the theory and, hopefully, stimulate our understanding of the semi-classical limit of loop quantum cosmology.


\section*{References}

\begin{thebibliography}{9}
\bibitem{Rev1}  Rovelli C 2004 {\em Quantum Gravity} (Cambridge: Cambridge University Press); Thiemann T 2007 {\em Introduction to Modern Canonical Quantum General Relativity} (Cambridge: Cambridge University Press)
\bibitem{lqc}   Ashtekar A and Singh P 2011 {\it Class.\ Quant.\ Grav.} {\bf 28} 213001; Banerjee K, Calcagni G and Mart\'{\i}n-Benito M 2011 {\it SIGMA} to appear (\arX{1109.6801})
\bibitem{BH1}   Bojowald M and Hossain G M 2007 {\it Class.\ Quant.\ Grav.} {\bf 24} 4801; 
2008 {\it Phys.\ Rev.} D {\bf 77} 023508 
\bibitem{CH}    Calcagni G and Hossain G M 2009 {\it Adv.\ Sci.\ Lett.} {\bf 2} 184 
\bibitem{BHKS}  Bojowald M, Hossain G M, Kagan M and Shankaranarayanan S 2008 {\it Phys.\ Rev.} D {\bf 78} 063547; 
 2009 {\it Phys.\ Rev.} D {\bf 79} 043505; 
  2010 {\it Phys.\ Rev.} D {\bf 82} 109903(E)
\bibitem{BC}    Bojowald M and Calcagni C 2011 {\it J.\ Cosmol.\ Astropart.\ Phys.} JCAP03(2011)032
\bibitem{APS}   Ashtekar A, Paw{\l}owski T and Singh P 2006 {\it Phys.\ Rev.} D {\bf 74} 084003 
\bibitem{BCT1}  Bojowald M, Calcagni G and Tsujikawa S 2011 {\it Phys.\ Rev.\ Lett.} {\bf 107} 211302
\bibitem{BCT2}  Bojowald M, Calcagni G and Tsujikawa S 2011 {\it J.\ Cosmol.\ Astropart.\ Phys.} JCAP11(2011)046
\bibitem{MCBG}  Mielczarek J, Cailleteau T, Barrau A and Grain J 2011 \arX{1106.3744}
\bibitem{Mie1}  Mielczarek J 2008 {\it J.\ Cosmol.\ Astropart.\ Phys.} JCAP11(2008)011; 
 Copeland E J, Mulryne D J, Nunes N J and Shaeri M 2009 {\it Phys.\ Rev.} D {\bf 79} 023508; 
  Grain J and Barrau A 2009 {\it Phys.\ Rev.\ Lett.} {\bf 102} 081301 
\end{thebibliography}
\end{document}